# Adaptive Safety for Internet of Things in e-Health

**Denis Trček[1], Habtamu Abie[2], and Åsmund Skomedal[2]**

[1] Faculty of Computer and Information Science, University of Ljubljana, Večna pot 113, 1000 Ljubljana, Slovenia; denis.trcek@fri.uni-lj.si

[2] Norwegian Computing Center, P.O.Box 114, Blindern, NO-0314 Oslo, Norway; {habtamu.abie, asmund.skomedal}@nr.no

**Abstract:** Wireless pervasive computing devices are rapidly penetrating our environments, where e-Health is among the most critical ones. In the e-Health environment it is not only important to address security and privacy issues (which are usually in the focus), but also safety needs for appropriate treatment. With an increasing proportion of pervasive (smart dust) devices which lack computing power, security, privacy, and safety provisioning in such environments is a demanding task - not to mention additional requirement about their adaptive provisioning. However, some twenty years ago an interesting research branch in computing domain appeared, called trust management. This branch is considered as a very handy alternative to support traditional hard security and privacy approaches. It is, therefore, often referred to as soft security (privacy) provisioning mechanism. As trust is inherently adaptive and as security and safety are much related, this paper presents a new approach where trust management is deployed in e-Health environments to enable adaptive safety provisioning. This paper also introduces a trustworthiness calculation framework that extends trust management methods with a comparative analysis of computational trust in Internet of Things (IoT).

**Keyword:** Internet of Things; safety; pervasive computing; e-health; trust management, adaptive safety provisioning.

## I. INTRODUCTION

Emerging smart objects technologies, often also collectively referred to as Internet of Things (IoT), are expected to provide numerous benefits in e-Health. However, these objects often lack computing resources, be it processing power, memory resources, communication capacity (data transfer rate), not to mention energy consumption, or related constraints.

Moreover, when such devices are used in the e-Health environments the above factors are clearly affecting safety features. Of similar concerns are privacy and security features. This paper intentionally focuses on non-adverse environments and is not addressing privacy and security. The reason is as follows: Security and privacy typically get more attention and belong to mainstream communications research, while pure functional, non-adverse environment operational issues are paid less attention to, although they are equally important. To properly provide non-life-threatening solutions in e-Health, the complete security-privacy-safety triple needs to be covered, rather than just security and privacy. It is worth to add that it is expected that safety solutions can serve as lessons learnt that can be applied later to security and privacy as well.

The research described in this paper is related to earlier work done in the ASSET project (http://asset.nr.no), which has focused on adaptive security (more details on these efforts can be found in e.g., [1-6]). The work in this paper extends these efforts by providing a new approach that uses computational trust management to ensure adaptive safety for IoT deployments in the e-Health environments, and trustworthiness calculation framework. As it is about safety, we particularly study variables like link quality, data transmission rate, response time, and uptime. In this setting sensor nodes (motes) constantly monitor one another and measure (observe) these parameters. Based on these measurements (observations) motes perform their trust assessments and deploy operators to make decisions about which motes to communicate with and how to perform communication.

The paper is structured as follows. In the second section computational trust management methods are briefly addressed to provide the basis for their deployment that is presented and discussed in the third section. Section four presents a framework for calculating trustworthiness of trust assessment. Section five presents a comparative analysis of related work. There are conclusions and future work in the sixth section. The paper ends with acknowledgements and references.

## II. Trust Management For Internet of Things

Computational trust management has now some two decades long history. According to [7] it can be divided into two epochs. The first epoch (that lasted approx. to the year 2000) was more about technologies that were trust enabling like public key infrastructure and web sites contents filtering. Only the second epoch that followed afterward was focused on trust as such. Now that the landscape of proposed trust management methods has exploded, only some more closely related to this paper will be presented.

One among most often cited trust management methods is Jøsang's Subjective logic [8]. It is based on the Dempster-Shafer theory of evidence (this is a theory that can be seen as a generalization of Bayesian statistics). The key function is the belief mass $m$ function, which is defined as

$$m: 2^\Theta \rightarrow [0,1],$$

where $m\{\ \} = 0$, and $\sum_{A \subseteq 2^\Theta} m(A) = 1$. Using this definition, the definition of three new functions, $b$ (belief), $d$ (disbelief) and $u$ (uncertainty) can be given:

$$b(X) = \sum_{Y \subseteq X} m(Y), \qquad d(X) = \sum_{X \cap Y = \emptyset} m(Y),$$

$$u(X) = 1 - (b(X) + d(X))$$

The triplet $\omega = (b, d, u)$ is referred to as opinion, which essentially represents trust in Jøsang's subjective logic. Using a particular agent's opinions new aggregates are defined. One such example is consensus.

Another suitable method is the Naïve trust management method [9]. It is designed for peer-to-peer architectures, and it directly deploys Bayesian reasoning – the well-known Bayesian formula is the basis:

$$p(H|D) = \frac{p(H,D)}{p(D)} = p(D\ |\ H) * \frac{p(H)}{p(D)},$$

In this formula $p(H)$ denotes the prior probability of $H$ before $D$ is observed, $p(D|H)$ denotes the probability of $D$ being observed when $H$ is true, and $p(D)$ denotes the unconditional probability of $D$. To obtain probabilities of safety-relevant questions like "What is the probability that an interaction will be trusted given that peer's transfer rate and uptime are considered?" The extended Bayesian formula can be used:

$$p(H|(D_1, D_2)) = \frac{p(H, D_1, D_2)}{p(D_1, D_2)} = \cdots = p(D_1|(H, D_2)) * \frac{p(H|D_2)}{p(D_1|D_2)}$$

The third suitable, but significantly different method is Qualitative Assessment Dynamics, QAD (see, e.g., [7] or [4]). It is of anthropocentric nature and can be used when human users evaluate motes (through, e.g., an artificial agent that runs on a certain mote). Therefore, QAD deploys operators that reflect humans reasoning when it comes to trust. Trust values in QAD (i.e., assessments denoted as $\alpha_{i,j}$ meaning agents $i$ trust toward agent $j$) can be fully trusted (denoted as 2), partially trusted (denoted as 1), undecided (denoted as 0), partially distrusted (denoted as -1) and fully distrusted (denoted as -2). In cases where an agent is not aware of another agent (or does not want to disclose its assessment) the trust assessment is marked by a dash "-".

Trust assessments in a society with $n$ agents can be conveniently represented by assessment matrix **A**. In such matrix values in a row state particular agent's assessments toward other agents in a community, while values in a given column state assessments of other agents toward a selected agent – the most general form of assessment matrix for a society with $n$ agents is as follows:

$$A = \begin{bmatrix} \alpha_{1,1} & \alpha_{1,2} & \cdots & \alpha_{1,n} \\ \alpha_{2,1} & \alpha_{2,2} & \cdots & \alpha_{2,n} \\ \vdots & \vdots & \ddots & \vdots \\ \alpha_{n,1} & \alpha_{n,2} & \cdots & \alpha_{n,n} \end{bmatrix}$$

Dynamics in a society (trust dynamics) is enabled by selected operators, which are functions $f_i$ such that, $f_i: A_{n,j} = (\alpha^-_{1,j}, \alpha^-_{2,j}, \alpha^-_{3,j}, \ldots, \alpha^-_{n,j}) \rightarrow \alpha^+_{i,j}$, $i = 1, 2, \ldots, n$, where "$i$" denotes the $i$-th agent, superscript "−" denotes pre-operation value, and superscript "+" post-operation value. This formula states that agent "$i$" takes into account all the assessments in column "$j$" and using appropriate operator computes the new assessment. Further, excluding undefined assessments from a trust vector, a society assessment sub-vector is obtained, denoted as $\boldsymbol{A}_{n1,k} = (\alpha_{1,k}, \alpha_{2,k}, \ldots, \alpha_{n1,k})$, where index "$n_1$" denotes non-undefined values in a society trust vector. It is now possible to define the following operators:

a) $\alpha^-_{i,j} \neq -$:

$d_i$:
$$\begin{cases} \alpha^-_{i,j} \rightarrow \alpha^+_{i,j} & \text{if } \frac{1}{n_1}\sum_{k=1}^{n_1} \alpha^-_{k,j} \leq \alpha^-_{i,j} \\ \alpha^-_{i,j} + 1 \rightarrow \alpha^+_{i,j} & \text{otherwise} \end{cases}$$

$g_i$:
$$\begin{cases} \alpha^-_{i,j} \rightarrow \alpha^+_{i,j} & \text{if } \frac{1}{n_1}\sum_{k=1}^{n_1} \alpha^-_{k,j} \geq \alpha^-_{i,j} \\ \alpha^-_{i,j} - 1 \rightarrow \alpha^+_{i,j} & \text{otherwise} \end{cases}$$

$k_i$:
$$\begin{cases} \left\lceil \frac{1}{n_1}\sum_{k=1}^{n_1} \alpha^-_{k,j} \right\rceil \rightarrow \alpha^+_{i,j} & \text{if } \frac{1}{n_1}\sum_{k=1}^{n_1} \alpha^-_{k,j} < 0 \\ \left\lfloor \frac{1}{n_1}\sum_{k=1}^{n_1} \alpha^-_{k,j} \right\rfloor \rightarrow \alpha^+_{i,j} & \text{otherwise} \end{cases}$$

$h_i$: $\quad random(-2, -1, 0, 1, 2) \rightarrow \alpha^+_{i,j}$

b) $\alpha^-_{i,j} = -$:

$$- \rightarrow \alpha^+_{i,j}$$

The first operator is called moderate optimistic operator, the second is moderate pessimistic operator, the third is centralistic consensus seeker operator, and the fourth is assessment hoping. These descriptive names of operators reflect the semantics behind the above stated corresponding definitions.

Naïve trust management, Subjective logic, and QAD are not the only possible candidates for practical deployment in adaptive safety settings. However, they cover some main approaches well and provide the basis for description and understanding of deployment of such methods.

## III. TRUST MANAGEMENT BASED ADAPTIVE SAFETY

### A. Adaptive Safety Framework

Adaptive systems can adjust their behavior in response to their perception of the environment and the systems themselves. They often use feedback loops to control their dynamic behavior [10,11,43]. These control loops are important features and, consequently, should be elevated to "first-class" structures in the modelling, design, and implementation of any adaptive systems [12]. Our framework, therefore, applies the monitor-analyze-adapt (plan, execute and learn) feedback mechanisms as depicted in Figure 1. The actual effective reaction on trust decisions is realized during the "adapt" phase using the actuators. The analysis phase of our approach uses different trust assessment frameworks as described in the previous section to analyze and assess the monitored safety parameters from the "monitor" phase, and to assess the trustworthiness of the assessed trust. The "monitor" phase gathers contextual safety parameters both from within the sensor nodes and from the environment using soft "sensors". Actuators here refer to the selectors of communication priorities and the choosers of safety path adaptively. The actual implementation of the different phases can be offloaded to either trusted sensor nodes or distributed among trusted sensor nodes. The deployment of this framework is described in detail in the ensuing section.

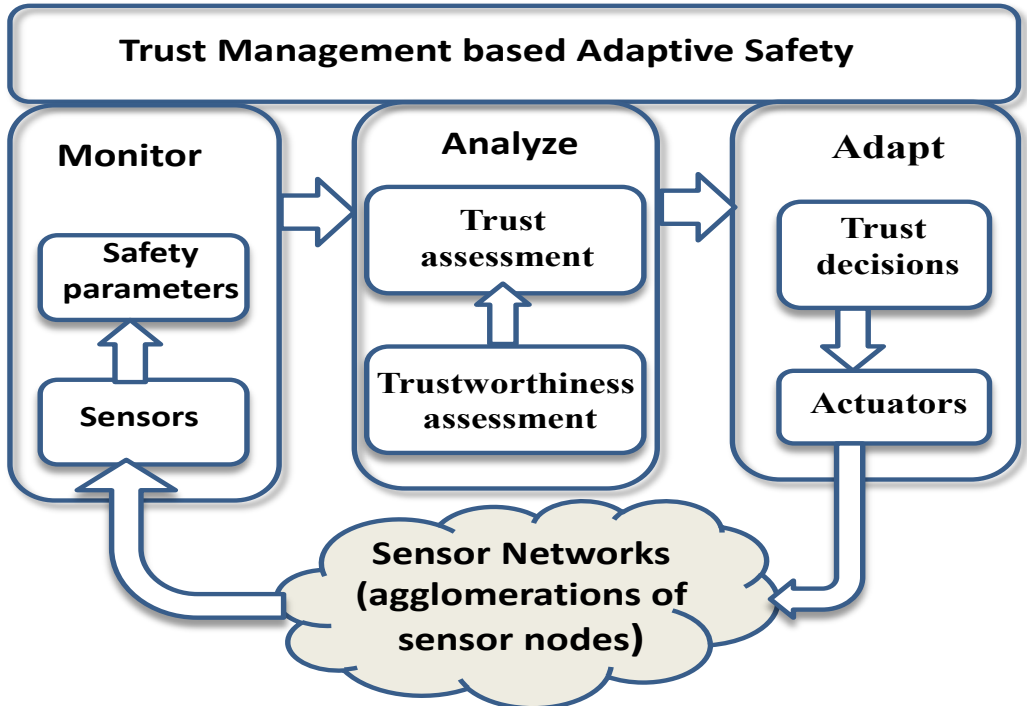

Figure 1: Framework for trust based adaptive safety.

## B. Deployment of trust management based adaptive safety

As stated earlier, the research problem of this paper is the provision of safety in non-malicious environments. Deploying trust management for adaptive safety means the following: Assume that a mote has to decide about priority of its communication and wants to ensure that the best possible path is chosen in terms of safety (i.e., operational parameters). To solve this problem, the mote bases its decision about with whom to communicate on deploying a particular trust management method.

In principle, this solution seems straightforward for implementation, but closer evaluation reveals quite some challenges. The problem of trust assessments calculation is computationally hard [13]. More precisely, having $n$ individual entities in a society, the number $N$ of trust relations that has to be taken into account is as given in the equation below (note that individuals and their aggregates with $m$ individuals, i.e., combinations of these individuals, have to be considered):

$$N = \left( \sum_{m=1}^{n} \binom{n}{m} \right)^2$$

This is still only the number of relations that has to be submitted to a trust-reasoning engine – and only when the operators are taken into account it is possible to obtain the final computational complexity.

Therefore, adaptive safety solutions should be able to handle this issue. This is however going directly against the main limitation of such environments – power and energy constraints. To the best of our knowledge, this is one among the first attempts to adaptively address the above problem, which has to be first conceptually well elaborated. We therefore start addressing the problem by considering two typical architectures:

- The first architecture consists of numerous sensor nodes that form agglomerations, where each of these agglomerations is tied to a particular sink node (or controlled by a corresponding agent that may reside on this node).
- The second architecture consists of independent agglomerations of sensor nodes that operate in a peer-to-peer manner within such agglomerations.

In the first architecture sink nodes are explicitly deployed and for them it can be assumed that there is no lack of computing resources. Consequently, an application of a particular trust management method is relatively straightforward. It can be done in a way where agents are considered to be the above-mentioned sink nodes, so actually trust of agglomeration vs. agglomeration is addressed. Within each agglomeration the corresponding sink node can deploy further trust management at sensors level.

The second architecture is more demanding, because sensor nodes (motes) operate as peers and they perform their tasks (e.g., collecting data, calculating aggregates, activating actuators, etc.) autonomously for an extended period of time. In this case with our assumption that motes are not malicious it makes sense to distribute trust management workload and perform adaptive soft safety as the following steps which will be referred to as Rolling Work-load Protocol (RWP):

1. Each mote starts a timer to monitor the duration of a globally agreed initial time interval ⊚$_i$.
2. As long as ⊚$_i$ does not expire, each mote polls its neighbors to evaluate safety parameters (e.g., link quality, data transmission rate, response time, and uptime) and forms neighborhood assessment (trust) matrix A$_{minor}$. Further, each mote monitors its residual energy together with energy renewal capacities, and calculates its projected (forecasted) computing capabilities (PCP) for the period ⊚$_i$. Each mote also calculates the local rate of change (RoC) in the neighborhood assessment matrix. The PCP and RoC can be expressed on a scale from 1 to 10 with 1 denoting the lowest computing capabilities and the smallest rate of change, while 10 denotes the highest computing capabilities and the largest rate of change.
3. Each mote floods the network with an energy efficient flooding algorithm (like, e.g., [14]) to announce its address and PCP value. Based on this flooding each mote finds the mote with the highest available computing power (HACP) mote. In case of more than one candidate some precedence is enforced, e.g. the mote is chosen with the lowest value of its network address.
4. After HACP mote is determined, it calculates matrix A$_{major}$ (trust matrix of the whole society) by receiving A$_{minor}$ matrices and operators from all motes. HACP also calculates the next time interval ⊚$_i$ a⊚$_{i+1}$ based on RoC values. A$_{major}$ and RoC are transmitted by using an energy aware routing algorithm like, e.g., LEACH [15] or PEGASIS [16] (an extensive overview of early such protocols can be found in [17] and [18], while the latest attempts are described in [19] and [20]).
5. During this new ⊚$_i$ interval, when motes need assessments about other motes, they send their requests to the selected HACP and obtain calculated assessment aggregates. Motes also perform tasks from step 2 during this step. After expiration of the new ⊚$_i$ step 3 is repeated.

RWP is, in principle, resistant to mote failures, because in the worst case only for the period ⊚$_n$ trust management support will not be available. If this is unacceptable, another option is to modify RWP in such a way that each mote selects two HACP motes – if the first one fails, the second gets contacted.

Clearly, many parameters come into play to optimize RWP, or derive other similar candidate protocols for adaptive safety purposes. However, this exceeds the scope of this paper.

## IV. FRAMEWORK FOR ASSESSING TRUSTWORTHINESS

This section presents a framework for assessing and calculating the trustworthiness of trust assessments $\alpha_{i,j}$. This framework uses trust and confidence assessments and maps them into trustworthiness assessment metric. The framework is based on an earlier work by Savola and Abie [21]. The definitions of trust, confidence, and trustworthiness are similar to those presented by Savola and Abie, with the following exceptions:

- trust $\alpha_{i,j}$ means trust assessment (belief) of an entity about another entity;
- confidence means the level of accuracy of and/or the assurance in the $\alpha_{i,j}$ trust relationship; and

- trustworthiness means the level of trust in the reliability of the $\alpha_{i,j}$ and a measurement of the degree to which the accuracy of and/or the assurance in this trust is trustworthy and can be verified.

The values of trust and confidence are expressed as numbers between zero and one, based on Bayesian probability. A trust value equal to one indicates absolute trust while a value close to zero indicates low trust. Similarly, a confidence value equal to one indicates high confidence in the accuracy of the trust value and a value close to zero indicates low confidence. Furthermore, a trustworthiness value close to zero indicates untrustworthiness and a value close to one indicates high or complete trustworthiness.

To calculate trustworthiness (and assuming normal distribution), some notations associated with $\alpha_{i,j}$ are defined as follows:

- $t(\alpha_{i,j})$: trust value of $\alpha_{i,j}$, having the property of $0 \leq t(\alpha_{i,j}) \leq 1$;
- $\sigma(\alpha_{i,j})$: standard deviation of trust value of $\alpha_{i,j}$; and
- $c(\alpha_{i,j})$: confidence value of $\alpha_{i,j}$, $0 \leq c(\alpha_{i,j}) \leq 1$.

Our framework uses the modified Bayesian approach to evaluate trust. So $\alpha_{i,j}$ is assumed to be trusted with probability $\theta$, while the Beta distribution is used for $\theta$ since it is the most promising due to its flexibility and simplicity and its conjugate is also a Beta distribution [21]. The trust value of the $\alpha_{i,j}$ can thus be calculated as the expectation value of the Beta distribution $Beta(\theta, A, B)$:

$$t(\alpha_{i,j}) = E(Beta(\theta, A, B)) = \frac{A}{A+B}$$

where A and B denote the degree of normal behaviors and misbehaviors, respectively. In this case, while normal behavior means that $\alpha_{i,j}$ can be trusted, misbehavior means that $\alpha_{i,j}$ cannot be trusted. The standard deviation of the trust value $t(\alpha_{i,j})$ is calculated as follows:

$$\sigma(\alpha_{i,j}) = \sigma(Beta(\theta, A, B))$$
$$= \sqrt{\frac{AB}{(A+B)^2(A+B+1)}}.$$

The confidence value of $\alpha_{i,j}$ is calculated as follows:

$$c(\alpha_{i,j}) = 1 - \sqrt{12\sigma(Beta(\theta, A, B))}$$
$$= 1 - \sqrt{\frac{12AB}{(A+B)^2(A+B+1)}}.$$

We can, for instance, define confidence ($c$) values of $0 \leq c < 0.2$ indicating no confidence, $0.2 \leq c < 0.5$ indicating low confidence, $0.5 \leq c < 0.8$ indicating good confidence, and values of $0.8 \leq c \leq 1.0$ indicating a high level of confidence. To facilitate trust-based decisions, trust and confidence are combined into a single value. This value, trustworthiness $T(\alpha_{i,j})$, is measured by combining the estimated levels of trust and confidence with some rules for the interpretation, and has the following property: $0 \leq T(\alpha_{i,j}) \leq 1$. If the confidence value of $\alpha_{i,j}$ is high, the trust value of $\alpha_{i,j}$ plays a more important role for trustworthiness. Thus, the trust of $\alpha_{i,j}$ has a larger weight than the confidence value of $\alpha_{i,j}$. Conversely, if the confidence value of $\alpha_{i,j}$ is low, the confidence value of $\alpha_{i,j}$ is clearly more important than the trust value of $\alpha_{i,j}$. Therefore, the trust value of $\alpha_{i,j}$ should have less weight than the confidence value of $\alpha_{i,j}$. The trustworthiness $T(\alpha_{i,j})$, associated with $t(\alpha_{i,j})$ and $c(\alpha_{i,j})$ is defined as:

$$T(\alpha_{i,j}) = 1 - \frac{\sqrt{\frac{(t-1)^2}{x^2} + \frac{(c-1)^2}{y^2}}}{\sqrt{\frac{1}{x^2} + \frac{1}{y^2}}}$$

where $x$ and $y$ are parameters that determine the relative importance of the trust value of $\alpha_{i,j}$., with $t$ denoting $t(\alpha_{i,j})$ versus the confidence value of $\alpha_{i,j}$, and $c$ denoting $c(\alpha_{i,j})$. Zouridaki et al. [22] have shown that the appropriate values of $x$ and $y$ are $\sqrt{2}$ and $\sqrt{9}$, respectively, for mapping trust and confidence to trustworthiness, and can be adjusted to the needs of a particular application. The trustworthiness values $T$ can also be defined as $0 \leq T < 0.2$ indicating a result that is not trustworthy, $0.2 \leq T < 0.5$ representing low trustworthiness, $0.5 \leq T < 0.8$ indicating good trustworthiness, and values of $0.8 \leq T \leq 1.0$ indicating high trustworthiness.

The assessment and calculation of the trustworthiness of the adaptive safety system as a whole can be achieved through the aggregation and propagation of trust assessment carried out in the system at different levels.

## V. RELATED WORK

The objective of the present work is to establish computational trust management based adaptive safety for IoT in eHealth. In addition to the related work mentioned in Section 2 and Section 3.B, this section provides a comparative review and analysis of research contributions in computational trust management, trustworthiness assessment, and safety for IoT.

### A. Research Contributions

**Trust management**

Trust management for Internet of Things (IoT) has been the focus of several surveys. In the area of IoT, Yan et al. [23] investigate the properties of trust, propose objectives of IoT trust management, and provide a survey on the current literature advances towards trustworthy IoT, and argue that IoT for reliable data fusion and mining, qualified services with context-awareness, and enhanced user privacy and information security, trust management plays an important role by helping people overcome perceptions of uncertainty and risk. In the area of Social Internet of Things (SIoT), a paradigm to integrate social networking into the IoT allowing people and devices to interact enabling a variety of attractive applications, authors in [24] survey basic concepts, properties and models proposed for the SIoT trust management, and argue that the establishment of a level of trustworthiness is essential for leveraging the interactions among things and social networks. In [25], existing trust management systems for online social communities are surveyed, four types of attacks are listed, and existing systems' vulnerabilities are analyzed. Compared to previous surveys, they take trust modeling, trust inference, and attacks into account. Authors in [26] discuss the concept and potential application areas of trust management for WSNs and IoT worlds, survey different trust management issues (i.e., cluster, aggregation, reputation), and suggest trust management should consider sensors topology, coverage deployment, target tracking, localization and IoT applications, and consider cluster and aggregation structure for creating reputation mechanism for the trust management in the IoT application area. In [27], three trust management models have been surveyed, namely PeerTrust model, subjective model and dynamic model, and a comparative study of these models have been conducted describing the pros and cons of each of them. In [28], a context-aware and multiservice trust management system for the IoT is presented, and heterogeneity in nodes, capabilities and services, and deters a class of common attacks designed to target trust management systems have been addressed.

However, all the surveyed work falls short of addressing a framework for building trust management based adaptive safety for the IoT with extended RWP protocol and trustworthiness assessment.

**Trust Computation**

Trust can help make decisions. However, it is a challenge to map trust into computational models because of its subjective properties, i.e., trust calculation is computationally hard [13]. In [29], existing IoT trust computation models are classified based on five design dimensions: trust composition, trust propagation, trust aggregation, trust update, and trust formation, and discuss the pros and cons of existing IoT trust computation models in terms of the class they fall within. For each class of IoT trust computation models, they discuss the defense mechanisms used and their effectiveness against malicious attacks and identified the most effective trust computation techniques as applying to IoT systems. Based on this classification, trust computation models for IoT systems for service management have been surveyed in [30]. In [31], a survey on and a comparison of various trust computing approaches targeted to MANETs is presented. The authors analyzed various works on trust dynamics including trust propagation, prediction and aggregation algorithms, the influence of network dynamics on trust dynamics, and the impact of trust on security services. In [32], a trust management model based on fuzzy reputation for IoT is proposed, considering a specific IoT with wireless sensors, and using QoS trust metrics like packet forwarding/delivery ratio and energy consumption. In [33], a dynamic trust management protocol for IoT systems is presented. It deals with misbehaving nodes whose behavior may change dynamically, with a formal treatment of the convergence, accuracy, and resilience properties of the protocol, and validate the effectiveness of the protocol with a trust-based service composition application in IoT environments. Their approach fits in well with our notion of trust dynamics (dynamics in a society) and RWP protocol. Furthermore, their contributions to trust evaluation in both static and dynamic environments gave clues on the effect of Alpha and Beta in our trustworthiness assessment framework. In [34], a formal trust management control mechanism is developed based on architecture modeling of IoT through decomposition of the IoT into three layers: sensor layer, core layer, and application layer. Each layer is controlled by trust management for special purpose: self-organized, affective routing and multi-service, respectively. The authors use a formal semantics-based and fuzzy set theory to realize the trust mechanism and argue that their result provides a general framework for the development of trust models for IoT.

**Trustworthiness Assessment**

In [35], a subjective model for the evaluation of trustworthiness in the SIoT is defined based on solutions proposed for P2P networks. In this model, each node computes the trustworthiness of its friends on the basis of its own experience and on the opinion of the common friends with the potential service providers. The model employs a feedback system and combines the credibility and centrality of the nodes to evaluate the trust level. In [36], this subjective model is extended with an objective model for trustworthiness management where the information about each node is distributed and stored making use of a distributed hash table structure. These two models fit in well with our notion of adaptive safety feedback loop to control the dynamic behavior of trust-based safety. Our model provides agents the ability to adjust their behavior in response to their perception of the environment and the systems themselves. Furthermore, our model incorporates a trustworthiness assessment framework using trust and confidence metrics. In [37], a general trust management framework is proposed for evaluating the trustworthiness of agents based on measurement theory using different trust factors. The framework uses two metrics – trustworthiness and confidence to represent trust. Similarly, a quantitative trust establishment framework is presented in [22] for reliable data packet delivery in MANETs (mobile ad hock networks) aiming to improve the reliability of packet forwarding over multi-hop routes. Their framework defines trust and confidence metrics which are then mapped to trustworthiness (opinion) metrics. These last two approaches are similar in part to our trustworthiness assessment framework in using trust and confidence to assess trustworthiness but differ in part in the calculation of trust and confidence and mapping them to trustworthiness in a dynamic environment and for safety in IoT.

**Trust Evaluation**

In [38], unified review criteria on trust evaluation in the IoT are defined and a review on trust evaluation for IoT by comparing several trust evaluation schemes is provided. The authors argue that there is still a lack of a fully suitable distributed and dynamic solution for the flexible and scalable IoT context despite a lot of research on trust management in the IoT. In [39], on the other hand trust evaluation model for the SIoT is presented, comprised of the triad of trust indicators reputation, experience and knowledge, and the authors argued that their model covers multi-dimensional aspects of trust by incorporating heterogeneous information from direct observation (as knowledge), personal experiences (as experience) to global opinions (as reputation). An evaluation model is also provided for these with an aggregation mechanism for deriving trust values as the final outcome of the evaluation model. In [27], the different trust metrics that are used for evaluating trust are analyzed and insights into how a trust metric can be customized to meet the requirements and goals of target IoT system are given. In [40], the importance of feedback is discussed. The discussion includes the different aspects to appropriately rate a service by generating consistent feedback for the trust evaluation in the exchange of services in IoT, how the concept of feedback has been exploited so far in trust management systems, and the different aspects of the feedback for multi-dimensional concept.

**Safety**

Pundits in the field, Bruce Schneier[i] and Ross Anderson[ii], have started blogging on safety and security concerns about the IoT. Leverett et al. [41] describe these concerns and outline the opportunities for governments, industry and researchers. They argue that European institutions and regulatory networks need cybersecurity expertise to support safety, privacy, consumer protection and competition, while their main message for industry and practitioners is that safety and security are merging: safety engineers are going to have to learn all about security, and vice versa, and researchers will have lots of new topics, from the design of the next generation of regulatory institutions to technical topics such as sustainability of software and the toolchains that support it. For all this to be sustainable, the trust which users place in them must be well-founded. For this to be sustainable, the trust which users place in them must be well-founded. Sfar et al. [42] argues that safety in IoT should focus on several operations such as control, command, surveillance, communications, intelligence, and reconnaissance, etc., meeting the need of intelligent objects, ensuring their safety across their whole life cycle, and improving persons' safety by reducing injuries and fatalities during the manufacturing process.

## B. Discussion

The unique characteristics and usage scenarios of IoT in eHealth introduce new safety challenges. The increasing proportion of pervasive (smart dust) devices which lack computing power, security, privacy, and safety provisioning in such environments is a challenge - not to mention provisioning of adaptivity to tackle dynamicity. An accurate and resilient trust-based adaptive safety assessment on trust level of IoT entities is required. Given the dynamics in a society (trust dynamics) the ability of agents to adjust their soft safety in response to their perception of the environment and the systems themselves should be provided. Although there are many research contributions about computational trust in IoT systems, most of them have not considered these and fall short defining a framework for building trust-based adaptive safety for IoT with rolling work-load protocol for independent agglomerations of sensor nodes that operate in a peer-to-peer manner, and trustworthiness assessment.

Table 1: Summary of computation trust

| References | Safety | Computation | Evaluation | Confidence | Trustworthiness | Dynamics | Adaptivity |
|---|---|---|---|---|---|---|---|
| [22] |  | + | + | + | + |  |  |
| [27] |  | + |  |  |  | + |  |
| [29, 30] |  | + | + |  |  |  |  |
| [32] |  | + | + |  |  |  |  |
| [33] |  | + | + | + | + | + |  |
| [34] |  | + | + |  |  | + |  |
| [35,36] |  | + | + |  |  | + |  |
| [37] |  |  |  | + | + |  |  |
| [40] |  | + | + |  |  | + | + |
| [41] | + |  |  |  |  |  |  |
| Proposed approach | + | + | + | + | + | + | + |

Table 1 summarizes the computation trust solutions which have been presented in the literature to address computation trust-based safety in the IoT environment. We compare these solutions based on their ability to capture safety, computational trust, trust evaluation, confidence and trustworthiness assessment, trust dynamics, and adaptivity.

## VI. CONCLUSION

We are becoming increasingly dependent on smart objects that range from NFCs, RFIDs with sensors and bare sensors on one side of the spectrum to capable sensor motes and smart phones on the other side of the spectrum. These structures are becoming deployed in various settings, among them being also e-Health settings. However, in this latter case not only security and privacy, but also safety is of prime concern with additional important tweak being the requirement for its adaptivity. As it is not a trivial task to provide adaptive behavior with traditional approaches, this paper deploys a new approach where trust management methods (which are inherently adaptive) are deployed. The paper also exposes the importance of safety, which is usually considered only second to security and privacy and presents a new protocol for use of existing trust management methods for this purpose in an energy-aware manner. The presented protocol, called RWP, builds on other existing lower level (and energy aware) protocols in order to fulfill the final goal – security, privacy and safety provisioning for advanced services in e-Health. In addition, the paper presents a framework for assessing and calculating the trustworthiness of assessments.

Our future work will include more rigorous algorithms for the calculation of the trustworthiness of the adaptive safety system as a whole considering many parameters which come into play to optimize RWP or derive other similar candidate protocols for adaptive safety purposes.

**Acknowledgments:** Authors acknowledge the support of the Slovenian Research Agency ARRS through research program Pervasive computing, P2-0359. In addition, this work is a result of the Norwegian project ASSET (Adaptive Security for Smart Internet of Things in e-Health) led by the Norwegian Computing Center and funded by the Research Council of Norway in the VERDIKT program, grant no. 2131310/O70, and the IoTSec (Security in IoT for Smart Grids) project, funded by the Research Council of Norway in the IKTPLUSS program, grant no. 248113/O70.

**References**

[1] Yared Berhanu Woldegeorgis, Habtamu Abie, and Mohamed Hamdi, "A Testbed for Adaptive Security for IoT in eHealth", *Proc. of Int. Workshop on Adaptive Security & Privacy Management for the Internet of Things (ASPI) '13*, Zürich, 2013.

[2] Reijo Savola and Habtamu Abie, "Metrics-Driven Security Objective Decomposition for an E-Health Application with Adaptive Security Management", *Proc. of the Int. Workshop on Adaptive Security & Privacy Management for the Internet of Things (ASPI) '13*, Zürich, 2013.

[3] Stefan Poslad, Mohamed Hamdi, and Habtamu Abie, "Adaptive Security & Privacy management for the Internet of Things (ASPI 2013)", *Proc. of UbiComp '13 Adjunct Proceedings of the 2013 ACM conference on Pervasive and ubiquitous computing adjunct publication*, New York, 2013, pp. 373-378.

[4] Trček D., Brodnik A., "Hard and soft security provisioning for computationally weak pervasive computing systems in e-health", *IEEE Wireless Communications*, vol. 20, no. 4, pp. 22-29, Aug. 2013.

[5] M. Hamdi, H. Abie, "Game-based adaptive security in the Internet of Things for eHealth", 2014 IEEE International Conference on Communications (ICC), 920-925, June 2014

[6] Wolfgang Leister, Mohamed Hamdi, Habtamu Abie, Stefan Poslad, Arild Torjusen, An Evaluation Framework for Adaptive Security for the IoT in eHealth, International Journal on Advances in Security Volume 7, Number 3 & 4, 2014, 93-109

[7] Trček D., "Trust management methodologies for the Web", *Reasoning Web: semantic technologies for the Web of data, LNCS 6848*, Springer, 2011, pp. 445-459.

[8] Jøsang A., "A logic for uncertain probabilities", *Int. Journal of Uncertainty, Fuzziness and Knowledge-Based Systems*, vol. 9, no. 3, pp. 279-311, 2001.

[9] Wang Y., Vassileva J., "Trust and Reputation Model in Peer-to-Peer Networks", *Proc. Of the 3rd Int. Conference on Peer-to-Peer Computing (P2P'03)*, pp. 150, Linkøping, 2003.

[10] Habtamu Abie, "Adaptive Security and Trust Management for Autonomic Message-Oriented Middleware", *IEEE 6th International Conference on Mobile Adhoc and Sensor Systems (MASS '09)*, 2009, pp. 810-817.

[11] Habtamu Abie and Ilangko Balasingham, "Risk-based adaptive security for smart IoT in eHealth". In *Proc. of the 7th Inter. Conference on Body Area Networks (BodyNets '12)*, 2012, pp 269-275.

[12] R. de Lemos et al., "Software Engineering for Self-Adaptive Systems: A Second Research Roadmap", R. de Lemos et al. (Eds.): Self-Adaptive Systems, LNCS 7475, 2013, pp. 1–32.

[13] Trček D., "Trust management in the pervasive computing era", *IEEE security & privacy*, vol. 9, no. 4, pp. 52-55, 2011.

[14] Hassanzadeh, A., Stoleru, R., C. Jianer, "Efficient flooding in Wireless Sensor Networks secured with neighborhood keys", *Proc. of 7th Int. Conf. on Wireless and Mobile Computing, Networking and Communications (WiMob'11)*, IEEE, 2011, pp. 119-126.

[15] W. Heinzelman, A. Chandrakasan, and H. Balakrishnan, "Energy-Efficient Communication Protocol for Wireless Microsensor Networks", *Proceedings of the Hawaii Conference on System Sciences*, 2000.

[16] Lindsey, S., Raghavendra, C., "PEGASIS: Power-Efficient gathering in sensor information systems", *Proceeding of IEEE Aerospace Conference*, vol.3, 2002, pp. 1125-1130.

[17] Al-Karaki, J., Kamal, A., "Routing techniques in wireless sensor networks: a survey", *IEEE Wireless Communications*, vol. 6, no. 11, 6–28, IEEE, 2004.

[18] Akkaya, K., Younis, M., "A survey of routing protocols in wireless sensor networks", *Ad Hoc Networks*, vol. 3, no. 3, pp. 325–349, Elsevier, 2005.

[19] Ehsan, S. and Hamdaoui, B., "Communications Surveys Tutorials, A Survey on Energy-Efficient Routing Techniques with QoS Assurances for Wireless Multimedia Sensor Networks", *IEEE Comms. Surves and Tutorials*, vol. 14, no. 2, pp. 265-278, IEEE, 2012.

[20] Rahman, M.A., Anwar, S., Pramanik, M.I., Rahman, M.F., "A survey on energy efficient routing techniques in Wireless Sensor Network", *Proc. of the 15th International Conference on Advanced Communication Technology '13*, IEEE, 2013, pp. 200-205.

[21] Reijo Savola and Habtamu Abie, "Development of Measurable Security for a Distributed Messaging System", *International Journal on Advances in Security*, vol. 2, no. 4, 2010, pp. 358-380.

[22] C. Zouridaki, B. L. Mark, M. Hejmo and R. K. Thomas, "A quantitative trust establishment framework for reliable data packet delivery in MANETs", ACM Workshop on Security of Ad Hoc and Sensor Networks (SASN '05), Alexandria, VA, Nov. 7, 2005.

[23] Zheng Yan, Peng Zhang, and Athanasios V. Vasilakos, A Survey on Trust Management for Internet of Things, Journal of Network and Computer Applications, 42, 2014, 120–134

[24] W. Abdelghani, C.A. Zayani, I. Amous, F. Sdes, "Trust management in social internet of things: a survey", Conference on e-Business e-Services and e-Society (I3E2016), 13-15 September 2016, 430-441

[25] Yefeng Ruan, Arjan Durresi, A survey of trust management systems for online social communities - Trust modeling, trust inference and attacks, Knowledge-Based Systems, v.106 n.C, 150-163, August 2016

[26] Kai-Di Chang and Jiann-Liang Chen, A survey of Trust Management in WSNs, Internet of Things and Future Internet, KSII Transactions on Internet and Information Systems, Vol. 6, N. 1, Jan 2012

[27] T. R. Thamburu, A.V. Vinitha, A Survey on Trust Management Models in Internet of Things Systems, International Journal of Advanced Research in Computer Science and Software Engineering, Volume 7, Issue 1, January 2017

[28] Yosra Ben Saied, Alexis Olivereau, Djamal Zeghlache, and Maryline Laurent, Trust management system design for the Internet of Things: A context-aware and multiservice approach, computers & security xxx (2013) 1-15

[29] Jia Guo and Ing-Ray Chen, A Classification of Trust Computation Models for Service-Oriented Internet of Things Systems, Services Computing (SCC) 2015 IEEE International Conference on, pp. 324-331, 2015

[30] Jia Guo, Ing-Ray Chen, and Jeffrey J.P. Tsai, A survey of trust computation models for service management in internet of things systems, Computer Communications, Volume 97, 1 January 2017, 1-14

[31] Kannan Govindan and Prasant Mohapatra, Trust computations and trust dynamics in mobile adhoc networks: A survey. IEEE Commun. Surv. Tutor. 14, 2 (2012), 279-298.

[32] Dong Chen et al., TRM-IoT: A Trust Management Model Based on Fuzzy Reputation for Internet of Things, Comput. Sci. Inf. Syst. 8, 2011, 1207-1228.

[33] Fenye Bao and Ing-Ray Chen, Dynamic trust management for internet of things applications. In Proceedings of the 2012 international workshop on Self-aware internet of things (Self-IoT '12), 2012, 1-6, ACM, New York, NY, USA

[34] Gu Lize, Wang Jingpei and Sun Bin, Trust management mechanism for Internet of Things, China Communications, Volume: 11, Issue: 2, Feb 2014, 148-156.

[35] Michele Nitti, Roberto Girau, Luigi Atzori, Antonio Iera, Giacomo Morabito, A subjective model for trustworthiness evaluation in the social Internet of Things, 2012 IEEE 23rd International Symposium on Personal Indoor and Mobile Radio Communications (PIMRC), 9-12 Sept. 2012.

[36] Michele Nitti, Roberto Girau, and Luigi Atzori, Trustworthiness Management in the Social Internet of Things. IEEE Trans. on Knowl. and Data Eng. 26, 5 (May 2014), 1253-1266.

[37] Yefeng Ruan, Arjan Durresi, and Lina Alfantoukh. Trust management framework for internet of things. In Proceedings of the 2016 IEEE 30th International Conference on Advanced Information Networking and Applications (AINA), 2016, 1013-1019.

[38] Pu Wang and Peng Zhang, A Review on Trust Evaluation for Internet of Things. In Proceedings of the 9th EAI International Conference on Mobile Multimedia Communications (MobiMedia '16). ICST (Institute for Computer Sciences, Social-Informatics and Telecommunications Engineering), ICST, 2016, 34-39, Brussels, Belgium, Belgium

[39] Nguyen Binh Truong, Hyunwoo Lee, Bob Askwith and Gyu Myoung Lee, Toward a Trust Evaluation Mechanism in the Social Internet of Things, Sensors 2017, 17, 1346, 1-24.

[40] Michele Nitti, Roberto Girau, Luigi Atzori, Virginia Pilloni, Trustworthiness management in the IoT: The importance of the feedback, 20th Conference on Innovations in Clouds, Internet and Networks (ICIN), 7-9 March 2017

[41] Eireann Leverett, Richard Clayton, and Ross Anderson, Standardisation and Certification of the `Internet of Things', May 22, 2017, https://www.cl.cam.ac.uk/~rja14/Papers/weis2017.pdf

[42] Arbia Riahi Sfar, Enrico Natalizio, Yacine Challal, Zied Chtourou, A roadmap for security challenges in the Internet of Things, Digital Communications and Networks, April 2017.

[43] Trček, D., Abie, H., Skomedal, Å., and Starc, I. (2010). Advanced framework for digital forensic technologies and procedures. *Journal of forensic sciences*, *55*(6), 1471-1480.

---

[i] https://www.schneier.com/blog/archives/2017/06/safety_and_secu.html

[ii] https://www.lightbluetouchpaper.org/2017/06/01/when-safety-and-security-become-one/